\documentclass[%
 reprint,
superscriptaddress,
amsmath,amssymb,
aps,
pra,
floatfix,
]{revtex4-2}

\usepackage{graphicx}
\usepackage{dcolumn}
\usepackage{bm}
\usepackage{hyperref}
\usepackage[utf8]{inputenc}
\usepackage{color}

\renewcommand{\vec}[1]{{\boldsymbol{\mathrm{#1}}}}

\newcommand{\degree}{\ensuremath{{^{\circ}}}}

\newcommand{\sech}{\ensuremath{\operatorname{sech}}}

\newcommand{\sinc}{\ensuremath{\operatorname{sinc}}}

\newcommand{\Ei}{\ensuremath{\operatorname{Ei}}}
\newcommand{\Erf}{\ensuremath{\operatorname{Erf}}}

\definecolor{darkred}{rgb}{0.81, 0.06, 0.13}

\newcommand{\revI}[1]{{\color{black}#1}}

\begin{document}

\preprint{APS/123-QED}

\title{Einstein-Podolsky-Rosen correlations in spontaneous parametric down-conversion:\\ Beyond the Gaussian approximation}

\author{A. G. da Costa Moura}
\email{alex.gutenberg@ict.ufvjm.edu.br}
\affiliation{Instituto de Ci\^{e}ncia e Tecnologia, Universidade Federal dos Vales do Jequitinhonha e Mucuri, Rodovia MGT 367 - km 583, N 5000, Alto da Jacuba, Diamantina, MG 39100-000, Brazil%
 }%
\author{C. H. Monken}%
 \email{monken@fisica.ufmg.br}
\affiliation{%
 Departamento de F\'{i}sica, ICEx, Universidade Federal de Minas Gerais.
Av. Antonio Carlos, 6627, Belo Horizonte, MG, 31270-901, Brazil}%

\date{\today}

\begin{abstract}
We present analytic expressions for the coincidence detection probability amplitudes of photon pairs generated by spontaneous parametric down-conversion in both momentum and position spaces, without using the Gaussian approximation and taking into account the effects of birefringence in the nonlinear crystal. We also present experimental data supporting our theoretical predictions, using Einstein-Podolsky-Rosen correlations as benchmarks, for 8 different pump beam configurations. 
\end{abstract}

\maketitle


\section{\label{intro}Introduction}
\revI{Spontaneous parametric down-conversion (SPDC) is a versatile and widely used tool in investigating fundamental quantum properties of correlated two-photon fields. Among these properties, nonclassical transverse momentum and transverse position correlations in two-photon states have been explored in many works with SPDC, in particular, the so-called Einstein-Podolsky-Rosen (EPR) paradox \cite{EPR1935}, first realized experimentally by Howell \textit{et al.} \cite{Howell2004} and by D'Angelo \textit{et al.} \cite{Dangelo2004}. 

Realizing the EPR paradox consists of preparing a quantum state of two spatially separated particles that allows one to infer with high precision either the position or the momentum of one of the particles (say, particle 1) without interacting with it, by measuring the position or the momentum of the other particle (particle 2). Since the measurement of $x_{2}$ or $p_{2}$ is a matter of choice, and $x$ and $p$ are incompatible variables and therefore subjected to the uncertainty relation $\Delta x\Delta p\ge \hbar/2$, EPR used the possibility to prepare such a state and the hypothesis of locality to suggest that Quantum Mechanics is an incomplete theory although it is correct in its statistical predictions. After a long debate, remarkable theoretical developments, and a long series of experiments \cite{Genovese2005}, the idea that local hidden-variable theories are ruled out is now common sense. Nevertheless, EPR-type correlations are interesting on their own \cite{Reid2009} and have been studied over the last two decades in two-photon states generated by SPDC \cite{Tasca2009,Walborn2011,Edgar2012,Moreau2012,Moreau2014,Schneeloch2018,Giese2018,Chen2019,Ndagano2020,Srivastav2022,Bhattacharjee2022,Bhattacharjee2022b,Patil2023}. 

As a rule, most works on EPR-type correlations in SPDC either rely on oversimplified models to describe the two-photon quantum state or do not present a theoretical model in both momentum and position representations to fit experimental data. In general, those simplified models do not include the effects of birefringence in the nonlinear crystals used in practice. Because of this, EPR correlations in SPDC have been analyzed only in the direction that is not affected by birefringence, that is, the direction normal to the plane defined by the crystallographic optic axis and the pump beam propagation direction \cite{daCostaMoura2010}. The commonly used Gaussian models to describe SPDC two-photon states have the advantage of simplifying calculations, but fail to correctly describe the full state propagation and do not include the effects of birefringence.  A detailed discussion of Gaussian approximations in SPDC can be found in ref. \cite{Gomez2012}, although in comparison with a simplified non-Gaussian model. 

Experimental conditions such as the pump laser beam focusing, the nonlinear crystal length, and birefringence can affect EPR correlations in a way that is not explained by previous models.  In this work, we present a more precise theoretical model for the SPDC two-photon state in both position and momentum representations and show how well it fits experimental data.

This paper is organized as follows. In Sec. II we explain the meaning of the Gaussian approximation in SPDC and present an accurate expression for the two-photon state generated by SPDC in both momentum and position representations without making use of that approximation. In Sec. III we describe EPR correlations in the two-photon states generated by SPDC and how they depend on experimental conditions. In Sec. IV we present an experiment where we measured EPR correlations in SPDC with eight different pump beam configurations, to validate our theoretical predictions. In Sec. V we present a brief discussion of the results and our conclusions.} 

\section{\label{pwf}The two-photon state generated by SPDC}

\revI{The basic SPDC process occurs when one photon from the laser pump beam of frequency $\omega_{p}$, usually in the ultra-violet spectral range, is converted into two photons of frequencies $\omega_{1}$ and $\omega_{2}$, such that $\omega_{1}+\omega_{2}=\omega_{p}$ (energy conservation) and $\vec{k}_{1}+\vec{k}_{2}\approx\vec{k}_{p}$ (phase match). Due to dispersion, the refractive index at $\omega_{p}$ is greater than it is at the lower frequencies $\omega_{1}$ and $\omega_{2}$, making phase match impossible in isotropic media. This problem is circumvented in birefringent nonlinear media, where dispersion can be compensated by birefringence. For example, in negative uniaxial crystals \cite{Yariv1984}, such as BBO, the pump beam polarized in the extraordinary direction and the down-converted beams polarized in the ordinary direction can be subjected to the same refractive indices, provided they propagate at appropriate directions. If the pump laser beam with extraordinary polarization propagates at an angle $\theta$ with respect to the optic axis direction inside the crystal, it is subjected to a refractive index \cite{daCostaMoura2010}
\begin{align}
\eta_{p}=\frac{n_{op}n_{ep}}{n^{2}_{op}\sin^{2}\theta+n^{2}_{ep}\cos^{2}\theta},
\end{align}
where $n_{op}$ and $n_{ep}$ are the ordinary and extraordinary refractive indices, respectively, at $\omega_{p}$. The collinear type I phase match condition is achieved when $\theta$ is such that, $\eta_{p}\omega_{p}=n_{o1}\omega_{1}+n_{o2}\omega_{2}$, where $n_{o1}$ and $n_{o2}$ are the ordinary refractive indices at frequencies $\omega_{1}$ and $\omega_{2}$, respectively.}

\begin{figure}
\begin{center}
\includegraphics{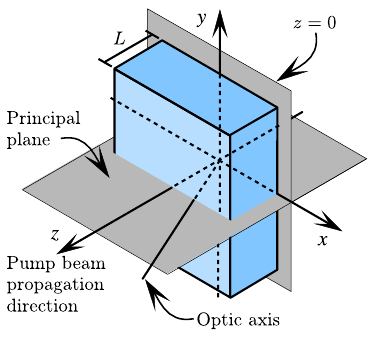}
\end{center}
\caption{\label{crystal} Nonlinear crystal configuration.}
\end{figure}

Let us consider a piece of negative birefringent nonlinear crystal (e.\,g., BBO) in the form of a block having its input face lying on the plane $z=0$, and cut for type I phase match with the principal plane (defined by optic axis and the pump beam propagation axis) parallel to the plane $xz$. A uv pump beam whose cross section lies entirely within the input and output faces of the crystal propagates along the $z$ axis with extraordinary ($x$) polarization. The crystal thickness in the $z$ direction is $L$. \revI{This configuration is illustrated in Fig. \ref{crystal}}.

In the k-vector (momentum) representation, the two-photon detection probability amplitude for the state generated by SPDC in the paraxial approximation is known to be well described (up to a normalization constant) by \cite{Walborn2010} 
\begin{align}
\psi(\vec{q}_{1},\vec{q}_{2})&=\mathcal{E}_{0}(\vec{q}_{1}+\vec{q}_{2})e^{-i\Delta_{oo}}\sinc \Delta_{oo} ,
\end{align}
where $\vec{q}_{j}$ is the $xy$ component of $\vec{k}_{j}$ ($j=1,2$), $\mathcal{E}_{0}(\vec{q})$ is the angular spectrum of the pump beam on the plane  $z=0$,
\begin{align}
\Delta_{oo}=\mu_{oo}+l_{t}(q_{1x}+q_{2x})
-\frac{L}{4k_{p}}\bigg|\sqrt{\frac{\omega_{2}}{\omega_{1}}}\vec{q}_{1}-\sqrt{\frac{\omega_{1}}{\omega_{2}}}\vec{q}_{2}\bigg|^{2},\nonumber
\end{align}
$k_{p}=\eta_{p}\omega_{p}/c$,  $l_{t}$ is half of the transverse walk-off length, $\mu_{oo}=(\bar{n}_{o}-\eta_{p})k_{p}L/2\eta_{p}$, and  $\bar{n}_{o}$ is the ordinary refractive index of the nonlinear crystal at $\omega_{p}/2$. The transverse walk-off length is given by \cite{daCostaMoura2010}
\begin{align}
2l_{t}&=\frac{(n_{op}^{2}-n_{ep}^{2})\sin\theta\cos\theta}{n_{op}^{2}\sin^{2}\theta+n_{ep}^{2}\cos^{2}\theta}L.
\label{lt}
\end{align}
\revI{In negative uniaxial crystals, $n_{ep}<n_{op}$ and the extraordinary beam tends do deviate away from the optic axis direction. This situation is illustrated in Fig. \ref{walkoff}
\begin{figure}
\begin{center}
\includegraphics{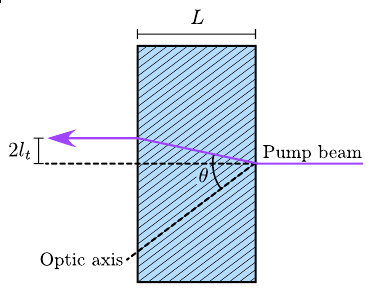}
\end{center}
\caption{\label{walkoff} The walk-off of the pump beam in a negative uniaxial crystal.}
\end{figure}
}

Here we work in the collinear phase match condition, where $\bar{n}_{o}=\eta_{p}\rightarrow\mu_{oo}=0$, and in the quasi-degenerate regime, where $\omega_{1}=(1+\nu)\omega_{p}/2$ and $\omega_{2}=(1-\nu)\omega_{p}/2$, with $\nu\ll 1$. This means that $\sqrt{\omega_{2}/\omega_{1}}\approx 1-\nu$ and  $\sqrt{\omega_{1}/\omega_{2}}\approx 1+\nu$. \revI{In the collinear phase match it is possible to make a straightforward use of the paraxial approximation and Fourier optics \cite{Goodman2005}. It is also possible to do so in the noncollinear phase match but at the cost of more complicated expressions. We chose to work in the quasi-degenerate regime in order to stress the generality of our model. The degenerate regime can be readily recovered by making $\nu=0$ in the expressions that follow.
It is interesting to notice that even in the collinear regime there will be a walk-off in the propagation of the pump beam inside the nonlinear crystal, since the phase match angle $\theta$ lies, in general, between $0$ and $\pi/2$. In our experiment (see section \ref{experiment}), $\theta\approx 33\degree$.}

To simplify the notation, we make
\begin{align}
\vec{Q}&=\vec{q}_{1}+\vec{q}_{2},\nonumber\\
\vec{P}&=(1-\nu)\vec{q}_{1}-(1+\nu)\vec{q}_{2},\nonumber\\
\beta ^{2} &=\frac{L}{4k_{p}}.
\end{align}
Then,
\begin{align}
\psi(\vec{Q},\vec{P})=\mathcal{E}_{0}(\vec{Q})\sinc(l_{t}Q_{x}-\beta ^{2}  P^{2})e^{-i(l_{t}Q_{x}-\beta ^{2}  P^{2})}.
\end{align}
\revI{Since $\psi(\vec{Q},\vec{P})$ is a two-beam plane-wave spectrum and the two beams propagate independently by acquiring a phase factor depending on the $z$-component of each k-vector \cite{Goodman2005}, on the output face of the nonlinear crystal, at $z=L$, the propagated $\psi(\vec{Q},\vec{P})$ is}
\begin{align}
\psi(\vec{Q},\vec{P},L)&=\psi(\vec{Q},\vec{P})e^{i(k_{1z}+k_{2z})L}.
\end{align}
Inside the crystal, in the collinear phase match \cite{Walborn2010},
\begin{align}
(k_{1z}+k_{2z})L&=\frac{\bar{n}_{o}L}{c}(\omega_{1}+\omega_{2})\nonumber\\
&-\frac{Lc}{\bar{n}_{o}}[(1-\nu)q_{1}^{2}+(1+\nu)q_{2}^{2}]\nonumber\\
&=k_{p}L-2\beta^{2}(Q^{2}+P^{2}).
\end{align}
Therefore,
\begin{align}
\psi(\vec{Q},\vec{P},L)&=\mathcal{E}_{0}(\vec{Q})\sinc(l_{t}Q_{x}-\beta ^{2}P^{2})\nonumber\\
&\times e^{-i(l_{t}Q_{x}+2\beta^{2}Q^{2})}e^{-i\beta^{2}P^{2}}. 
\end{align}

Considering a Gaussian pump beam, we can write, in the complex notation \cite{Siegman1986},
\begin{align}
\mathcal{E}(\vec{Q},z)=A\,e^{ik^{v}_{p}z}e^{-ia(z)Q^{2}/2k^{v}_{p}},
\end{align}
where $A$ is a constant,  $k^{v}_{p}=\omega_{p}/c$ (superscript $v$ stands for vacuum), $a(z)=z-z_{c}-ik^{v}_{p}w_{0}^{2}/2$, $w_{0}$ is the beam waist and $z_{c}$ is the waist location on the $z$ axis.
Hence,
\begin{align}
\mathcal{E}_{0}(\vec{Q})&=A\,e^{-ia_{0}Q^{2}/2k^{v}_{p}},
\end{align}
where $a_{0}=a(0)$. Then,  at the output face of the crystal,
\begin{align}
\psi(\vec{Q},\vec{P},L)&=e^{-i(a_{0}/2k^{v}_{p}+2\beta^{2})Q^{2}}e^{-il_{t}Q_{x}}e^{-i\beta^{2}P^{2}}\nonumber\\
&\times \sinc(l_{t}Q_{x}-\beta ^{2}P^{2}).
\end{align}
From $z=L$ to $z>L$, $\psi(\vec{Q},\vec{P})$ propagates in free space, that is,
\begin{align}
\psi(\vec{Q},\vec{P},z)=\psi(\vec{Q},\vec{P},L)e^{i(k^{v}_{1z}+k^{v}_{2z})(z-L)}.
\end{align}
In free space and collinear propagation,
\begin{align}
(k^{v}_{1z}+k^{v}_{2z})(z-L)&=\Big(k^{v}_{1}+k^{v}_{2}-\frac{q_{1}^{2}}{2k^{v}_{1}}-\frac{q_{2}^{2}}{2k^{v}_{2}}\Big)(z-L)\nonumber\\
&=k^{v}_{p}(z-L)+2\bar{n}_{o}\beta^{2}(Q^{2}+P^{2})\nonumber\\
&-\frac{z}{2k^{v}_{p}}(Q^{2}+P^{2}).
\end{align}
Therefore,
\begin{align}
\label{psiQPexact}
\psi(\vec{Q},\vec{P},z)&=\sinc(l_{t}Q_{x}-\beta ^{2}P^{2})e^{-il_{t}Q_{x}}\nonumber\\
&\times e^{-ib_{1}(z)Q^{2}}e^{-ib_{2}(z)P^{2}},
\end{align}
where
\begin{align}
b_{1}(z)&=a(z)/2k^{v}_{p}-2(\bar{n}_{o}-1)\beta^{2},\nonumber\\
b_{2}(z)&=z/2k^{v}_{p}-(2\bar{n}_{o}-1)\beta^{2}.
\end{align}

\revI{To the best of our knowledge, an accurate analytic expression for $\psi$ in the coordinate representation is not available in the literature, except under the Gaussian approximation, which consists of replacing $\sinc(l_{t}Q_{x}-\beta ^{2}P^{2})$ in Eq. \eqref{psiQPexact}  by a Gaussian function, necessarily neglecting the walk-off term $l_{t}Q_{x}$. This approximation works well when $L/k_{p}w_{0}^{2}\ll 1$, that is, when the pump beam is highly collimated, or when the crystal is very thin ($L\approx 1\,\mathrm{mm}$) \cite{Gomez2012}.}

To arrive at an expression for $\psi$, in coordinate representation, we adopt the following approximation: For pump beams with reasonably narrow-band angular spectra ($w_{0} \ge 50\,\mu$m) and $L\sim$ 1\,mm to\ 5\,mm, which cover most practical cases, we can make
\begin{align}
\label{psiQP}
\psi(\vec{Q},\vec{P},z)&=\sinc(l_{t}Q_{x})e^{-il_{t}Q_{x}}e^{-ib_{1}(z)Q^{2}}\nonumber\\
&\times \sinc(\beta ^{2}P^{2})e^{-ib_{2}(z)P^{2}},
\end{align}
which turns $\psi(\vec{Q},\vec{P},z)$ into a separable function of $\vec{Q}$ and $\vec{P}$.

Defining the coordinates $\vec{R}=[(1+\nu)\vec{\rho}_{1}+(1-\nu)\vec{\rho}_{2}]/2$ and $\vec{S}=(\vec{\rho}_{1}-\vec{\rho}_{2})/2$, the calculation of the Fourier transform of $\psi(\vec{Q},\vec{P},z)$ is straightforward:
\begin{align}
\label{psiRS}
\psi(\vec{R},\vec{S},z)&= \frac{e^{-R_{y}^{2}/4ib_{1}(z)}}{l_{t}\sqrt{ib_{1}(z)}}\nonumber\\
&\times\Big\{\Erf\Big[\frac{R_{x}-2l_{t}}{2\sqrt{ib_{1}(z)}}\Big]-\Erf\Big[\frac{R_{x}}{2\sqrt{ib_{1}(z)}}\Big] \Big\}\nonumber\\
&\times\Big\{\Ei\Big[\frac{ik_{p}^{v}\,S^{2}}{2(z-L)}\Big]-\Ei\Big[\frac{ik_{p}^{v}\,S^{2}}{2(z-L^{\prime})}\Big]\Big\}
\end{align}
for $z>L$, where $L^{\prime}=(1-1/\bar{n}_{o})L$, $\Erf$ is the error function, and $\Ei$ is the exponential integral function \cite{Lebedev1972}.

\revI{Eqs. \eqref{psiQP} and \eqref{psiRS} can be considered the main contribution of this work to the field of SPDC. Using Eq. \eqref{psiRS} one can predict experimental results of spatial two-photon correlations with very good accuracy and push experimental conditions to their limits when testing specific theoretical models. In this work, we test EPR correlations, as they involve both spatial and momentum correlations.}

\section{EPR correlations}
\begin{figure}[htb]
\begin{center}
\includegraphics{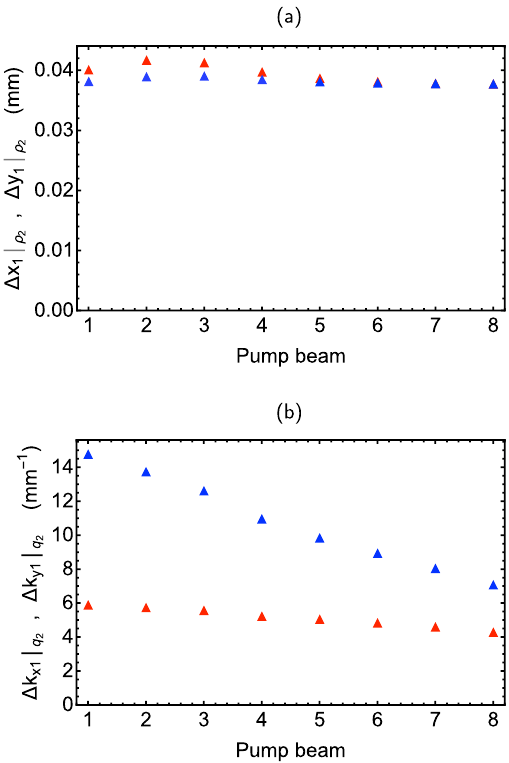}
\end{center}
\caption{\label{fig2} Predicted values of (a) $\Delta x_{1}|_{\vec{\rho}_{2}=0}$ (red) and $\Delta y_{1}|_{\vec{\rho}_{2}=0}$ (blue), (b) $\Delta k_{x1}|_{\vec{q}_{2}=0}$ (red) and  $\Delta k_{y1}|_{\vec{q}_{2}=0}$ (blue) for a 5\,mm-long BBO crystal pumped by 355\,nm laser beams whose parameters are listed in Table \ref{table1}.}
\end{figure}

With expressions \eqref{psiQP} and \eqref{psiRS} we can calculate the uncertainties $\Delta x_{1}$, $\Delta y_{1}$, for fixed $x_{2}$, $y_{2}$ and $z$, and $\Delta k_{x1}$, $\Delta k_{y1}$ for fixed $k_{x2}$, $k_{y2}$ as functions of the pump beam angular spectrum width (defined by the beam waist $w_{0}$) and see how they are affected by the crystal anisotropy (quantified here by walk-off parameter $l_{t}$). Assuming any fixed value for $\vec{\rho}_{2}$, Eq. \eqref{psiRS} allows us to calculate $\Delta x_{1}$ and $\Delta y_{1}$ for any $z>L$. Alternativelly, assuming any fixed value for $\vec{q}_{2}$,  Eq. \eqref{psiQP} allows us to calculate $\Delta k_{1x}$ and $\Delta k_{1y}$, which do not depend on $z$. As an example, Fig. \ref{fig2} shows plots of $\Delta x_{1}$ and $\Delta y_{1}$  for $\vec{\rho}_{2}=0$ and plots of $\Delta k_{x1}$ and $\Delta k_{y1}$ for $\vec{q}_{2}=0$, for a 5\,mm-long BBO crystal pumped by 355\,nm gaussian laser beams whose waists $w_{0}$ and waist positions $z_{c}$ are shown in Table \ref{table1}. Uncertainties $\Delta x_{1}$ and $\Delta y_{1}$ where calculated on the plane $z=5.0001$\,mm.

\begin{table}
\caption{\label{table1} Pump beam parameters.}
\begin{ruledtabular}
\begin{tabular}{c c c}
Beam & $w_{0}$ (mm) & $z_{c}$ (mm)\\ \hline
1 & 0.062 & 178\\
2 & 0.067 & 213\\
3 & 0.072 & 251\\
4 & 0.085 & 298\\ 
5 & 0.095 & 355\\
6 & 0.105 & 422\\
7 & 0.120 & 510\\
8 & 0.142 & 635
\end{tabular}
\end{ruledtabular}
\end{table}

One can see that while the position uncertainties in the $x$ and $y$ directions are about the same, the momentum uncertainties are strongly affected by the anisotropy. This is a direct consequence of the partial transfer of the pump beam angular spectrum in the direction parallel to the principal plane \cite{daCostaMoura2010}, in our case, the $x$ direction.

\begin{figure}[hbt]
\begin{center}
\includegraphics{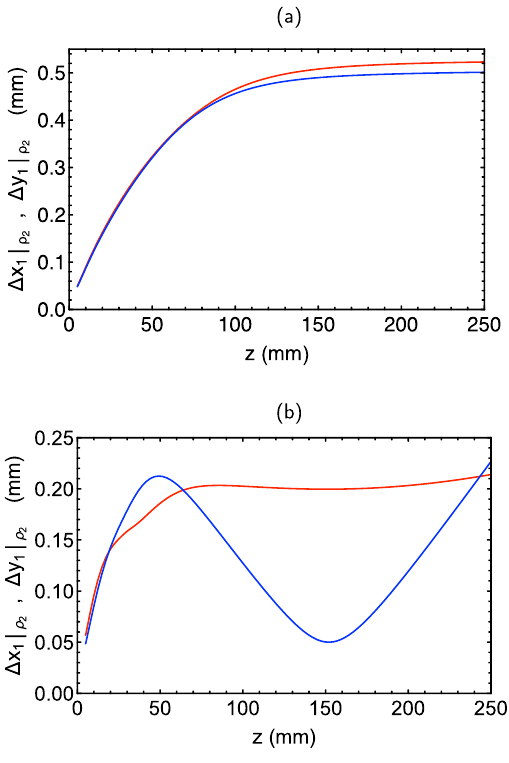}
\end{center}
\caption{\label{dxdyz} Predicted values of $\Delta x_{1}|_{\vec{\rho}_{2}=0}$ (red) and $\Delta y_{1}|_{\vec{\rho}_{2}=0}$ (blue) as functions of the distance from the output face of a 5\,mm-long BBO crystal cut for collinear degenerate phase match, pumped by a 355\, nm laser beam with (a) $w_{0}=0.5\,\mathrm{mm}$, $z_{c}=0$ and (b) $w_{0}=0.05\,\mathrm{mm}$, $z_{c}=150\,\mathrm{mm}$.}
\end{figure}

It is interesting to note that the uncertainties $\Delta x_{1}$ and $\Delta y_{1}$  increase rapidly as the distance from the crystal increases in the $z$ direction. In general, position uncertainties depend on the pump beam parameters,  as exemplified in Fig. \ref{dxdyz}. For a weakly focused laser beam ($w_{0}=0.5\,\mathrm{mm}$) whose waist is located at the crystal input face ($z=0$), our predictions shown in Fig. \ref{dxdyz}a are in good agreement with the results reported in Ref \cite{Bhattacharjee2022b}. This agreement is because in this case a Gaussian model fits well the uncertainty in position coordinates \cite{Gomez2012}.

\section{Experiment}
\label{experiment}
\begin{figure}[htb]
\begin{center}
\includegraphics{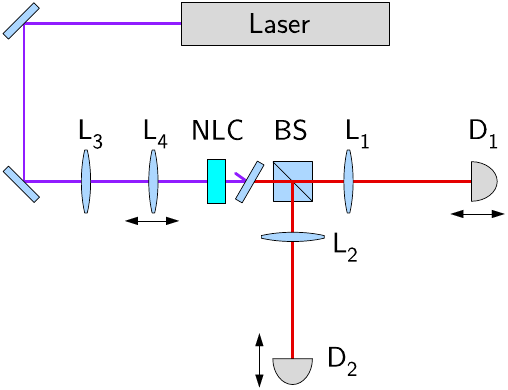}
\end{center}
\caption{\label{setup} Experimental setup}
\end{figure}

EPR correlations predicted in the previous section were tested experimentally with the setup represented in Fig. \ref{setup}. A 5\,mm-long BBO crystal cut for type I collinear phase match (\textit{NLC}) having its optic axis parallel to the $xz$ plane and input face located on the plane $z=0$ was pumped by a 355\,nm laser beam polarized in the $x$ direction, propagating along the $z$ direction. The beam parameters, listed in Table \ref{table1}, were changed with the help of a telescope composed by a lens $L_{3}$ of focal length 50\,mm and a lens $L_{4}$ of focal length 40\,mm separated from $L_{3}$ by a variable distance. The down-converted light, with $\lambda_{1}=690\,\mathrm{nm}$ and $\lambda_{2}=731\,\mathrm{nm}$ was sent to a beam splitter (\textit{BS}) and directed to detectors $D_{1}$ (equipped with a 12\,nm band-pass filter centered at 690\,nm) and $D_{2}$ (equipped with a 40\,nm band-pass filter centered at 730\,nm). Lenses $L_{1}$ and $L_{2}$ of focal length 75\,mm were placed at 150\,mm from the output face of the nonlinear crystal. Detectors $D_{1}$ and $D_{2}$ were placed at 75\,mm from $L_{1}$ and $L_{2}$ (Fourier plane) for the measurements of $|\psi(\vec{q}_{1},0,L)|^{2}$ and at 150\,mm (1:1 image plane) for the measurements of $|\psi(\vec{\rho}_{1},0,L)|^{2}$. Each detector consists of a multimode optical fiber with a diameter of $50\,\mu\mathrm{m}$, one tip mounted on a computer-controlled $xy$ motorized translation stage and the other tip coupled to a photon-counting avalanche photodiode. In all measurements, $D_{2}$ was kept at $\vec{\rho}_{2}=0$. 

\begin{figure}[htb]
\begin{center}
\includegraphics{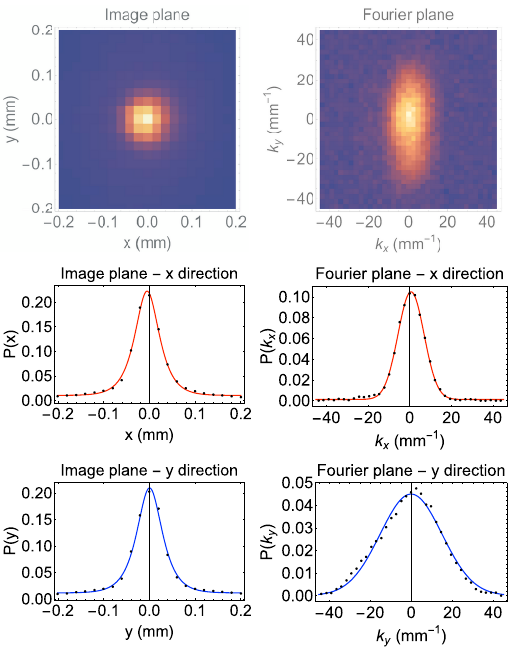} 
\end{center}
\caption{\label{examples} Top row: Examples of coincidence detection profiles on the image plane (left) and on the Fourier plane (right) for beam \# 1 (see Table \ref{table1}). Mid row: Corresponding  detection probability densities $P(x_{1}|\vec{\rho}_{2}=0)$ (left) and $P(y_{1}|\vec{\rho}_{2}=0)$ (right). Bottom row: Corresponding  detection probability densities $P(k_{x1}|\vec{q}_{2}=0)$ (left) and $P(k_{y1}|\vec{q}_{2}=0)$ (right). Dots are normalized experimental data and solid lines are best fits for distributions $A \sech \pi\xi/2\Delta_{\xi}$ ($\xi=x,y$ on the image plane) and $A \exp(-\xi^{2}/2\Delta_{\xi}^{2})$ ($\xi=k_{x},k_{y}$ on the Fourier plane). }
\end{figure}

\begin{figure}[htb]
\begin{center}
\includegraphics{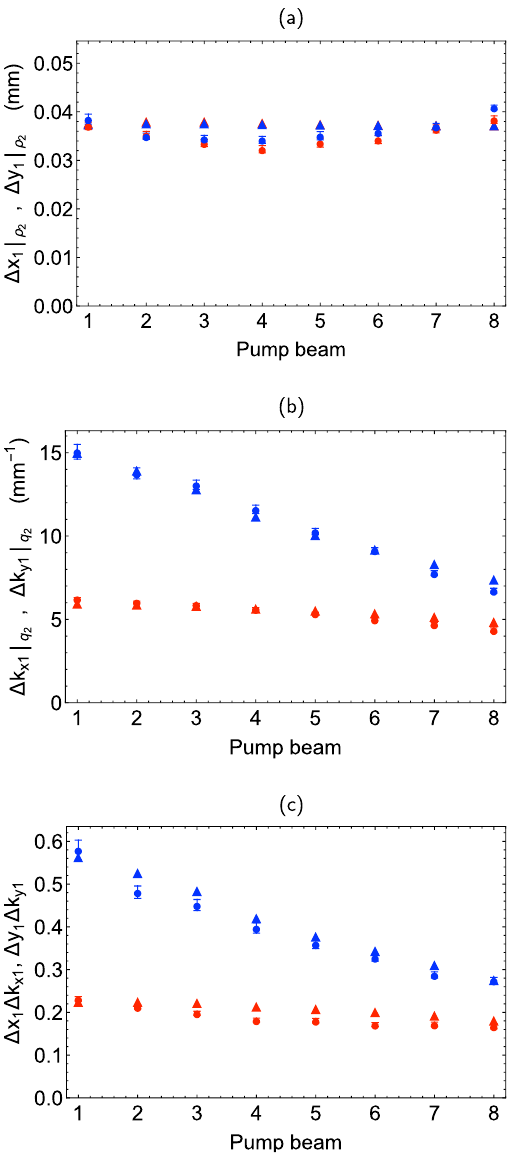}
\end{center}
\caption{\label{results} Measured ($\bullet$) and predicted ($\blacktriangle$) values of (a) $\Delta x_{1}$ (red) and $\Delta y_{1}$ (blue), (b) $\Delta k_{x1}$ (red) and  $\Delta k_{y1}$ (blue), (c) $\Delta x_{1}\Delta k_{x1}$ (red) and  $\Delta y_{1}\Delta k_{y1}$ (blue) for a 5\,mm-long BBO crystal pumped by 355\,nm laser beams whose parameters are listed in Table \ref{table1}.}
\end{figure}

Due to the relatively large fiber diameter and photon-counting fluctuations, experimental results are not directly comparable with theory.  To check the accuracy of theoretical predictions, the following procedure was adopted: \revI{Experimental data for coincidence detections on the image plane were fit to a hyperbolic secant distribution $A \sech \pi\xi/2\Delta_{\xi}$ ($\xi=x,y$), whose standard deviation is given by $\Delta_{\xi}$.} Numerical convolutions of the theoretical coincidence profiles with the 50\,$\mu\mathrm{m}$ circular apertures of $D_{1}$ and $D_{2}$ were made, resulting in the expected detection probability distributions. Profiles obtained when the crystal was pumped with the laser beam \#1 (see Table I) are shown in Fig. \ref{examples}. An additional correction of the beam waist radius was made, due to the pump beam $M^{2}$ factor of 1.14.  On the Fourier plane, a Gaussian distribution \revI{$A \exp(-\xi^{2}/2\Delta_{\xi}^{2})$} ($\xi=k_{x},k_{y}$) was used in a similar procedure.  Experimental results and theoretical predictions are presented in Fig. \ref{results}.

\section{\label{conclusion} Discussion and Conclusion}

From the results presented here, one can see that the EPR correlations $\Delta x\Delta k_{x}$ and $\Delta y\Delta k_{y}$ of the photons pairs generated by spontaneous parametric down-conversion are strongly affected by the pump beam angular spectrum in the direction normal to the principal plane (defined by the optic axis and the $z$ axis). Such dependence is much smaller in the direction parallel to the principal plane. This effect is readily explained by the presence of the term $\sinc l_{t}Q_{x}$ in Eq. \eqref{psiQP}. That term,  which depends on the birefringence, the phase match angle, and the crystal length (see Eq. \eqref{lt}), acts as a spatial filter for the transfer of the angular spectrum from the pump beam to the two-photon state \cite{daCostaMoura2010}. Because of this filtering effect, the product $\Delta x\Delta k_{x}$ behaves like the laser beam was less focused. The two uncertainty products  $\Delta x\Delta k_{x}$ and $\Delta y\Delta k_{y}$ tend to a unique minimum value as the pump beam gets more collimated, that is, $w_{0}\gg l_{t}$. In  our case, $l_{t}=0.186\,\mathrm{mm}$.

In conclusion, we have presented\revI{, unlike any previous publication}: (a) Accurate analytic expressions for the coincidence detection probability amplitudes of photon pairs generated by spontaneous parametric down-conversion in both momentum and position spaces \revI{on the entire plane normal to the pump beam}. Those expressions allow us to predict how the correlations in position and momentum depend on the system parameters like crystal length, crystal birefringence, pump beam focusing, pump beam waist location, and detectors locations. (b) Experimental data supporting our theoretical predictions, using Einstein-Podolsky-Rosen correlations as benchmarks, for 8 different pump beam configurations.

The results presented here may be useful in any application relying on position and momentum correlations of photon pairs generated by SPDC in birefringent crystals.

\acknowledgments
This work was supported by  CNPq Project 302872/2019-1 and INCT-IQ Project 465469/2014-0.

\end{document}